\documentclass[aps,prl,reprint,amsmath,amssymb,showpacs,longbibliography]{revtex4-1}

\usepackage{bm}
\usepackage{graphicx}
\usepackage{xcolor}
\usepackage{url}
\usepackage[normalem]{ulem}
\usepackage{times}
\usepackage{bbm}
\usepackage{mathrsfs} 
\usepackage{comment}
\usepackage{dsfont}
\usepackage{pdfpages}

\usepackage[colorlinks=true,breaklinks=true,allcolors=blue]{hyperref}
\usepackage{url}

\usepackage[normalem]{ulem}

\definecolor{michael}{rgb}{0.7,.3,.5}

\definecolor{daniel}{rgb}{0.2,.8,.2}

\newcommand{\pt}{\partial}

\renewcommand{\t}[1]{\tilde{#1}}

\newcommand{\NN}{\mathscr{N}}

\renewcommand{\H}{\mathscr{H}}
\newcommand{\kt}[1]{| #1 \rangle} 
\newcommand{\brkt}[2]{\langle #1 | #2 \rangle} 
\newcommand{\bropkt}[3]{\langle #1 |#2| #3 \rangle} 

\renewcommand{\a}{\alpha}

\newcommand{\s}{\sigma}

\renewcommand{\s}{\sigma}

\newcommand{\om}{\omega}

\renewcommand{\t}{\tau}

\newcommand{\sn}[2]{
\ensuremath{%
\left \{%
\begin{tabular}{c}%
$#1$ \\%
$#2$ %
\end{tabular}%
\right \}%
}%
}

\makeatletter
\let\cat@comma@active\@empty
\makeatother

\makeatletter
\AtBeginDocument{\let\LS@rot\@undefined}
\makeatother

\begin{document}

\title{\mbox{Classical Lieb-Robinson Bound for Estimating Equilibration Timescales of Isolated Quantum Systems}}

\author{Daniel Nickelsen}
\email{danielnickelsen@sun.ac.za}
\author{Michael Kastner}
\email{kastner@sun.ac.za}
\affiliation{National Institute for Theoretical Physics (NITheP), Stellenbosch 7600, South Africa}
\affiliation{Institute of Theoretical Physics,  Department of Physics, University of Stellenbosch, Stellenbosch 7600, South Africa}

\date{\today}

\begin{abstract}
We study equilibration of an isolated quantum system by mapping it onto a network of classical oscillators in Hilbert space. By choosing a suitable basis for this mapping, the degree of locality of the quantum system reflects in the sparseness of the network. We derive a Lieb-Robinson bound on the speed of propagation across the classical network, which allows us to estimate the timescale at which the quantum system equilibrates. The bound contains a parameter that quantifies the degree of locality of the Hamiltonian and the observable. Locality was disregarded in earlier studies of equilibration times, and is believed to be a key ingredient for making contact with the majority of physically realistic models. The more local the Hamiltonian and observables, the longer the equilibration timescale predicted by the bound.
\end{abstract}

\maketitle

Equilibration is one of the key concepts in thermodynamics. In the quest to derive, or at least justify, the macroscopic laws of thermodynamics from microscopic theories, much progress has been made on the quantum mechanical side over the last decade or two. For a variety of settings, rigorous proofs have been given, establishing conditions under which an isolated quantum mechanical system on a sufficiently large Hilbert space will approach equilibrium \cite{vonNeumann29,Tasaki98,Linden_etal09,Goldstein_etal10,ShortFarrelly12,Reimann10,KastnerReimann12,GogolinEisert16,Mori_etal18}. Key for the progress in the field was to identify suitable definitions of equilibration in a probabilistic sense: it is neither realistic to expect nor required that the density operator of the system converges to an equilibrium state; instead, equilibration happens on the level of observables, in the sense that expectation values of a suitable class of observables, including the physically realistic ones, approach their equilibrium values and stay close to them for most later times.

The aforementioned results are an important step towards a microscopic justification of thermodynamics. However, for explaining why equilibration and equilibrium are so ubiquitously observed in nature, one would need to show not only that equilibration takes place but also that it does so on a physically realistic timescale: neither too long for equilibrium ever to be attained nor too short for the equilibration process to be observed. The quest to provide a microscopic justification of physically realistic equilibration timescales is arguably the most important open question in the field \cite{deOliveira_etal18,Wilming_etal}. A key characteristic that determines whether a timescale can be considered realistic is its scaling with the Hilbert space dimension, because timescales that grow or decrease with the full dimension of the Hilbert space will result in either unrealistically long or short equilibration times \cite{GarciaPintos_etal17}.

General upper bounds on the equilibration timescale have been obtained \cite{ShortFarrelly12}, but the predicted timescales are unrealistically long. In fact, for a given quantum system, one can construct observables that equilibrate only after extremely long times \cite{GoldsteinHaraTasaki13,Malabarba_etal14}. Such behavior, however, is untypical in a well-defined probabilistic sense. On the other hand, it has also been shown that typical observables and/or Hamiltonians \cite{VinayakZnidaric12,Brandao_etal12,GoldsteinHaraTasaki13,Malabarba_etal14,Reimann16}, or typical nonequilibrium initial states \cite{GoldsteinHaraTasaki14}, equilibrate on unrealistically short timescales. It was conjectured that physically relevant systems are not typical in the sense of random matrix theory, and that {\em locality}\/ of observables and Hamiltonians needs to be taken into account in order to derive realistic equilibration timescales \cite{GoldsteinHaraTasaki13}.

An example of a local Hamiltonian $H$ is a lattice model with only nearest-neighbor interactions, and an example of a local observable $O$ is the total magnetization of a spin lattice model, given as the sum (over the entire lattice) of single-site spin operators. The simultaneous locality of $H$ and $O$ implies that there is a basis with respect to which the matrix representations of the two operators are simultaneously sparse. Similarly, by a theorem of Arad {\em et al.}\ \cite{AradKuwaharaLandau16}, $H$ is approximately sparse (up to exponentially small corrections) in the eigenbasis of the $O$, and {\em vice versa}.

\begin{figure}[t]
	\includegraphics[width=0.49\linewidth]{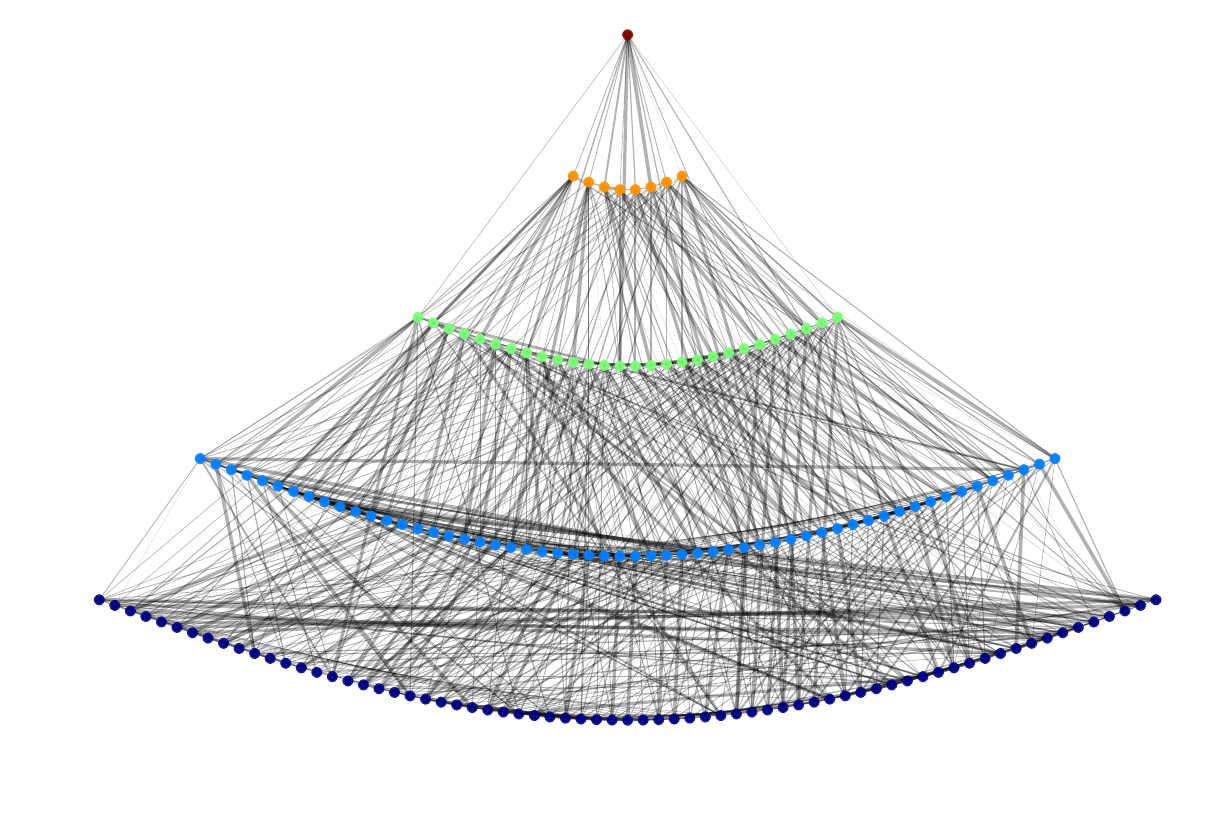}
	\includegraphics[width=0.49\linewidth]{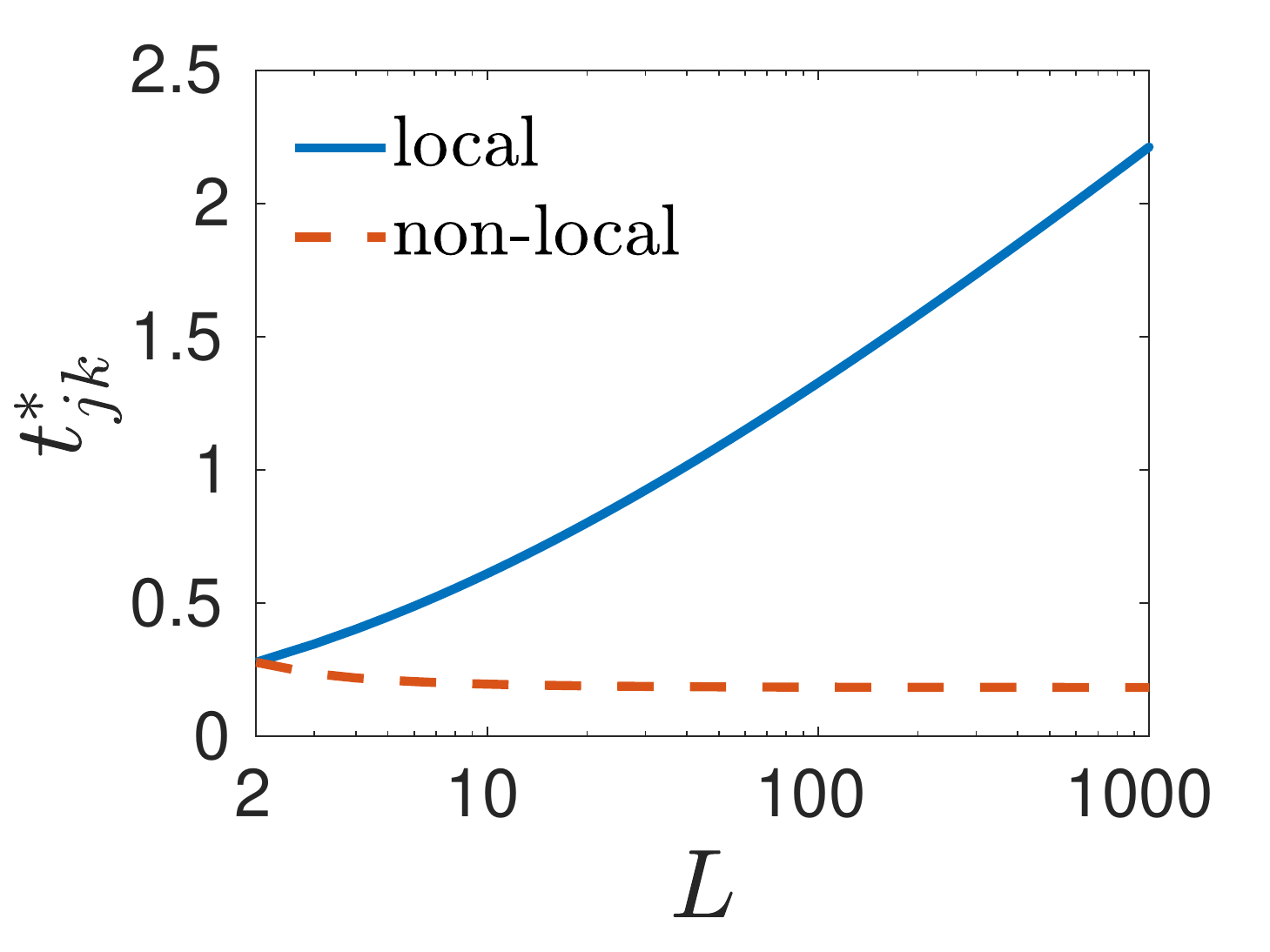}
	\caption{Left: Visualization of a typical weighted graph $H_{jk}$ for a local Hamiltonian. The top nodes (red) correspond to nonequilibrium states, and the bottom nodes (blue) are equilibrium states.
	The thickness of the edges is determined by $|H_{jk}|$.
	Right: Scaling of the equilibration timescale $t^\ast_{jk}$ from Eq.~\eqref{eq:tstar} with system size $L$ for particularly local and nonlocal choices of the parameters $g=\|H\|/L$ and $g=\|H\|(L-1)/L$, respectively. \label{fig:graph_tstL}}
\end{figure}

In this Letter, we make use of this sparseness to derive an estimate of the equilibration times of isolated, local quantum systems. The key idea is to rewrite the quantum mechanical time evolution governed by the Schr\"odinger equation as a network of coupled classical oscillators in Hilbert space, in which each node of the oscillator represents an eigenstate of the observable $O$, as illustrated in Fig~\ref{fig:graph_tstL} (left). Equilibration of the quantum system occurs when an excitation of an oscillator corresponding to a nonequilibrium value of $O$ propagates to an equilibrium node. We estimate the propagation speed through the network by means of a classical Lieb-Robinson bound, which in turn gives access to the equilibration time of the quantum system. Our main result is a bound on the equilibration time that, for sufficiently local interactions, scales logarithmically with the system size, and hence doubly logarithmic with the Hilbert space dimension, as shown in Fig~\ref{fig:graph_tstL} (right). Unlike, and complementary to, previously published upper bounds, our Lieb-Robinson approach provides a {\em lower} bound on the equilibration time. Moreover, the bound increases with increasing locality of the Hamiltonian and observable, leading to physically more realistic estimates.

{\em Setting.---}We consider a quantum system on the sites $\NN$ of a finite lattice or graph of size $L=|\NN|$, with a finite-dimensional Hilbert space $\H_i$ attached to each site $i\in\NN$. The total Hilbert space $\H$ is the tensor product space of all the $\H_i$. The dynamics of an isolated quantum system is generated by a time-independent Hamiltonian $H = \sum_X h_X$ acting on $\H$, where the summation is over subsets $X$ of $\NN$, and $h_X$ acts nontrivially only on the sites in $X$. Following Arad {\em et al.}\ \cite{AradKuwaharaLandau16}, we quantify the locality of $H$ through the parameter
\begin{align}\label{eq:g}
	g = \max_{i\in\NN} \Big\|\sum_{X\ni i}  h_X\Big\|.
\end{align}
Note that this rather weak definition of locality does not restrict interactions to only neighboring sites. We consider a $\kappa$-local observable
\begin{equation}\label{eq:O}
O=\sum_{X\in\NN_\kappa} \omega_X,
\end{equation}
i.e., the summation in \eqref{eq:O} extends only over the subsets $\NN_\kappa=\{X\in\NN : |X|\leqslant\kappa\}$ containing, at most, $\kappa$ sites, and $\omega_X$ acts nontrivially only on the sites in $X$. Without loss of generality we assume $\|O\|=1$.

According to the Schr\"odinger equation, the time evolution of a normalized initial state $\kt{\psi(0)}\in\H$ is given by $\kt{\psi(t)}=\exp(-iHt)\kt{\psi(0)}$. The longtime average of an observable $O$ can be written as
\begin{equation}
\langle O\rangle_\text{eq} \equiv \lim_{t\to\infty}\frac{1}{t}\!\int_0^t \! \bropkt{\psi(\t)}{ O}{\psi(\t)} d\t = \sum_k |c_k|^2 \bropkt{E_k}{O}{E_k},
\end{equation}
where $c_k=\brkt{E_k}{\psi(0)}$ are the overlaps of the initial state with the energy eigenstates $\kt{E_k}$. We define $\H_\text{eq}$ as the subspace of $\H$ spanned by those eigenstates $\kt{O_k}$ of $O$ for which the eigenvalues satisfy $|O_k-\langle O\rangle_\text{eq}|\leqslant\epsilon$ for some small positive $\epsilon$. Following Goldstein {\em et al.}\ \cite{GoldsteinHaraTasaki13}, we define equilibrium with respect to the observable $O$ as all the states in $\H$ which are sufficiently close to $\H_\text{eq}$. A nonequilibrium subspace $\H_\text{neq}$ can be defined in an analogous way and, from typicality arguments, it follows that $\dim\H_\text{eq}\gg\dim\H_\text{neq}$ \cite{GoldsteinHaraTasaki13}. According to these definitions, for a system to be in equilibrium with respect to the observable $O$ it is not sufficient that $\langle O\rangle\approx\langle O\rangle_\text{eq}$, but it is additionally required that the variance be small, $\langle (O-\langle O\rangle_\text{eq})^2\rangle\approx0$.

To analyze the dynamics that drives the system from non\-equi\-lib\-ri\-um to equilibrium, we choose a representation in the eigenbasis $\{\kt{O_k}\}$ of the observable $O$. With the definitions
\begin{align}
	x_j(t) = \brkt{O_j}{\psi(t)},\qquad 
	H_{jk} = \bropkt{O_j}{ H}{O_k},
\end{align}
we integrate the Schr\"odinger equation $\dot{x}_j(t) = -i\sum_k H_{jk} x_k(t)$ to obtain
\begin{align}\label{eq:SG_int_compo}
e^{iH_{jj}t}x_j(t) = x_j(0) - i\sum_{k\neq j} H_{jk} \int_0^t e^{iH_{jj}\tau} x_k(\tau) d\tau.
\end{align}
We interpret this equation as a network of oscillators $x_j(t)$, where the diagonal matrix elements $H_{jj}$ fix the frequencies, and the off-diagonal elements $H_{jk}$ determine the couplings between the oscillators.

The picture of coupled oscillators conveys an intuitive understanding of equilibration: Preparing the system in an initial state where oscillators corresponding to equilibrium observable eigenstates have negligible amplitudes $x_j(t)$, oscillations need to travel through the network in order to excite equilibrium oscillators. The locality of the quantum system imposes a locality structure on such a network of classical oscillators. This is in line with Refs.~\cite{AradKuwaharaLandau16,deOliveira_etal18}, where it was shown that a local observable is a banded matrix in the energy eigenbasis of a local Hamiltonian when the eigenstates are sorted in ascending order. An analogous result holds for the matrix representation of the Hamiltonian in the eigenbasis of the observable, which implies an approximate locality structure of the oscillator network, with coupling strengths $H_{jk}$ being zero or exponentially suppressed if $|O_k-O_j|$ is large. Figure~\ref{fig:graph_tstL} (left) illustrates that in order for the nonequilibrium state to excite the equilibrium states, excitations first need to travel through intermediate states.

{\em Bounds on propagation.---}The speed at which oscillations propagate through the network is therefore crucial for determining the timescale on which the quantum system equilibrates. To study the propagation speed we define
\begin{align}\label{eq:LR_obj}
	\varLambda_{jk}(t) \equiv \bigg|\frac{\pt x_j(t)}{\pt x_k(0)}\bigg|,
\end{align}
which quantifies the effect of a perturbation of the amplitude $x_k$ at time $0$ on the amplitude $x_j$ at a later time $t$. Upper bounds $B_{jk} \geqslant \varLambda_{jk}$ are known for fairly general Hamiltonian systems as classical analogs of Lieb-Robinson bounds \cite{Marchioro_etal78,Butta_etal07,RazSims09,IslambekovSimsTeschl12,MetivierBachelardKastner14}. For network nodes $j$ and $k$ separated by a large graph distance, $\varLambda_{jk}$ is small at early times, but will usually become non-negligible at later times. This onset of non-negligible $\varLambda_{jk}$ values gives rise to a causal structure in the plane of time $t$ and the graph distance on the network. Whether this causal structure has the shape of a light cone, or a generalization thereof, depends on the locality of the couplings \cite{MetivierBachelardKastner14}. In this way, the locality of the quantum system enters into our analysis.

Here we derive, by different techniques, a bound on $\varLambda_{jk}$ not as a function of the graph distance, but of the distance $|O_j-O_k|$ in observable eigenvalues, which is related to the distance from equilibrium if $\kt{O_j}$ is an equilibrium state. The time evolution equation $x_j(t)=\sum_k(e^{-iHt})_{jk}x_k(0)$ allows us to rewrite and upper bound Eq.~\eqref{eq:LR_obj} as
\begin{align}\label{eq:LR_obj2}
	\varLambda_{jk}(t) = \Big|\big(e^{-iHt}\big)_{jk}\Big| 
	\leqslant \sum_{n=0}^\infty \frac{t^n}{n!} \big|(H^n)_{jk}\big|.
\end{align}
We further bound the right-hand side of Eq.~\eqref{eq:LR_obj2} by combinatorial techniques, which are detailed towards the end of this Letter and in the Supplemental Material, to obtain our main result
\begin{widetext}
 \begin{align}
 	\ln\varLambda_{jk}(t) \leqslant B_{jk}(t) \equiv %
 	\begin{cases}
 		\displaystyle\left(\|H\| - \kappa gr\right)t - \tfrac{1}{2}|O_k-O_j|\left(\ln\frac{|O_k-O_j|}{2\kappa grt}-1\right)  & \text{for $\displaystyle t\leqslant t_{jk}^\star$}\\
 		0 & \text{for $t>t_{jk}^\star$}
 	\end{cases} %
 	\label{eq:LR_bd3}
 \end{align}
\end{widetext}
with
\begin{align}\label{eq:tstar}
	t_{jk}^\star = \tfrac{1}{2}|O_k-O_j|\frac{W\left(\frac{\|H\|-\kappa g r}{e \kappa g r}\right)}{\|H\| - \kappa gr}.
\end{align}
Here, $r\equiv\sum_X \| \om_X\|$ is an upper bound on the norm of the $\kappa$-local observable $O$ defined in Eq.~\eqref{eq:O}, and $W$ denotes the Lambert function defined via $W(ze^z)=z$ \cite{NIST}. 

\begin{figure}[b]
	\includegraphics[width=0.21\textwidth]{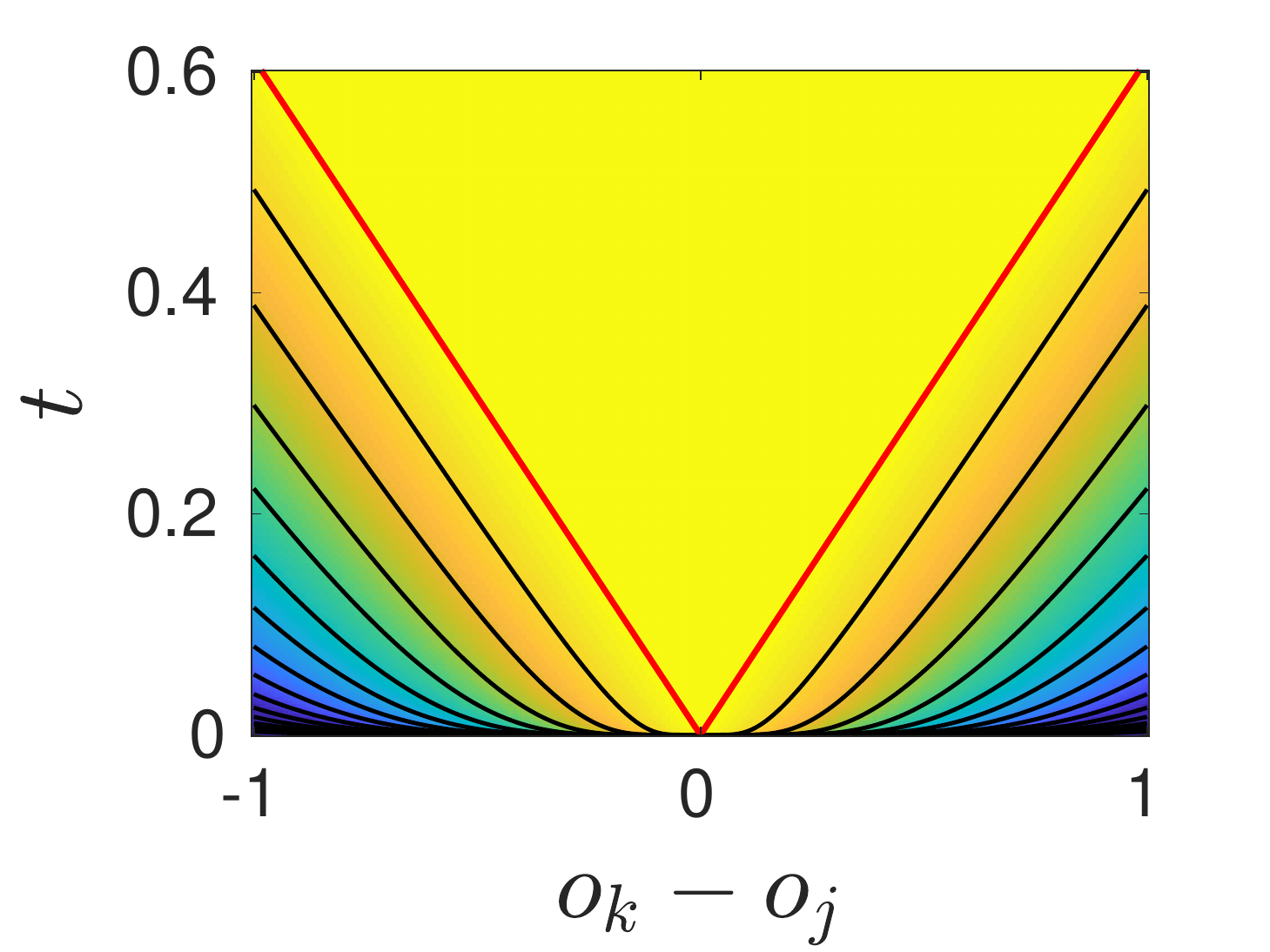}%
	\hspace{-7pt}
	\includegraphics[width=0.202\textwidth]{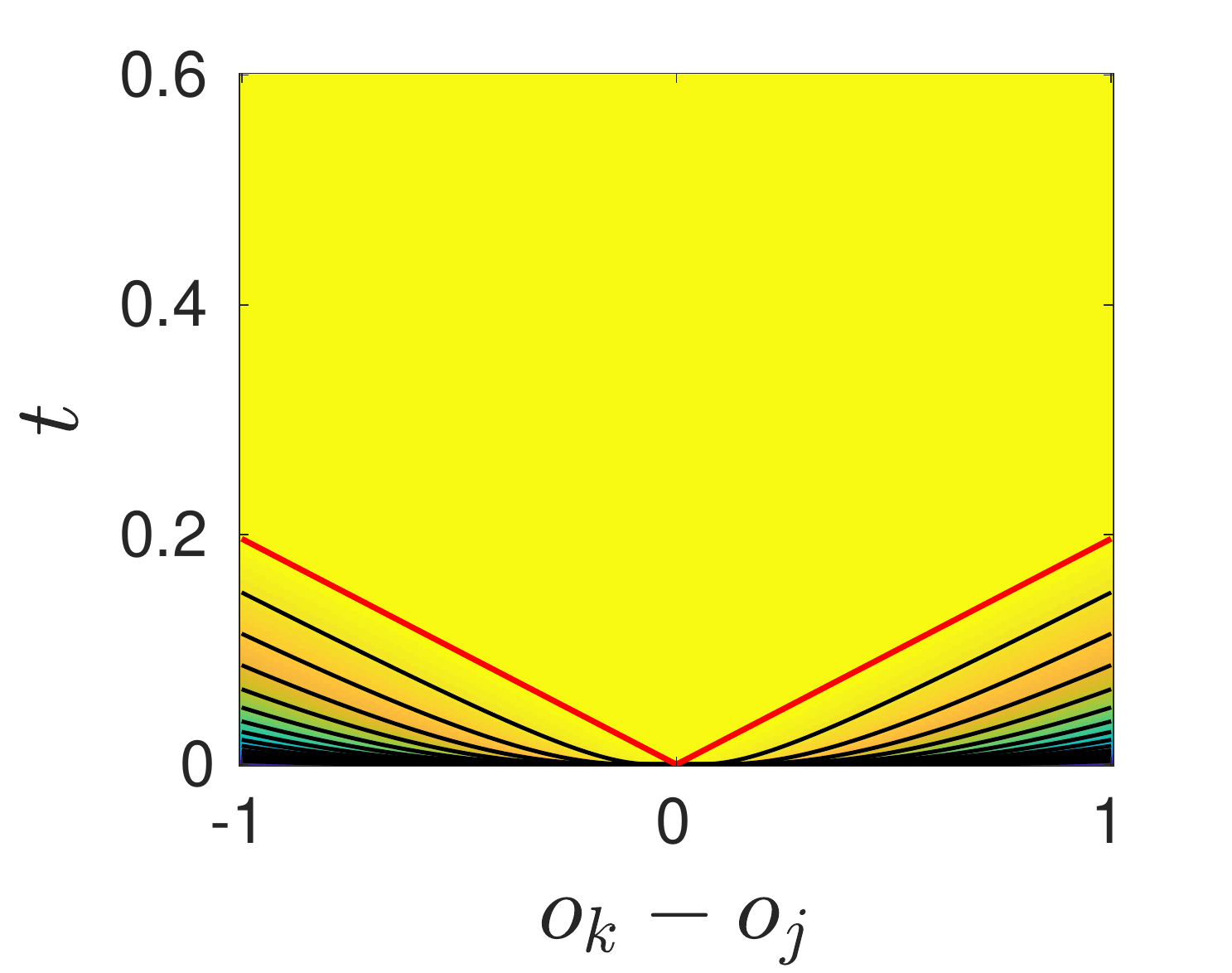}%
	\hspace{1pt}
	\includegraphics[trim=380pt 0 0 0,clip,height=0.121\textheight]{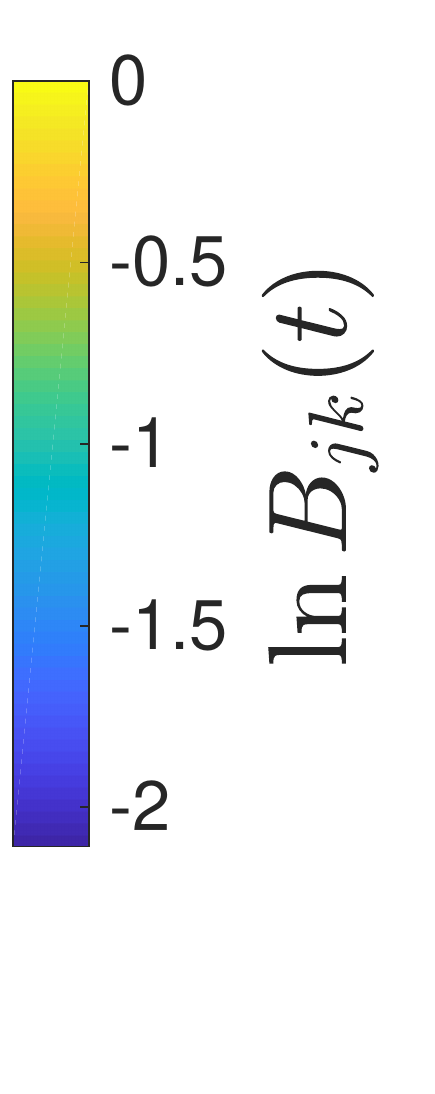}
	\caption{Contour plots of Lieb-Robinson bound $B_{jk}(t)$ in Eq.~\eqref{eq:LR_bd3} for parameter values $L=10$ and $\|H\|=\|O\|=k=r=1$. We use $g=\|H\|/L$ (left) for a strongly local system and $g=\|H\|(L-1)/L$ (right) for a nonlocal system. The time $t^\ast_{jk}$ in Eq.~\eqref{eq:tstar}, which defines a light cone, is shown as a solid red line. \label{fig:LR}}
\end{figure}

\begin{figure}[b]
	\includegraphics[width=0.50\linewidth]{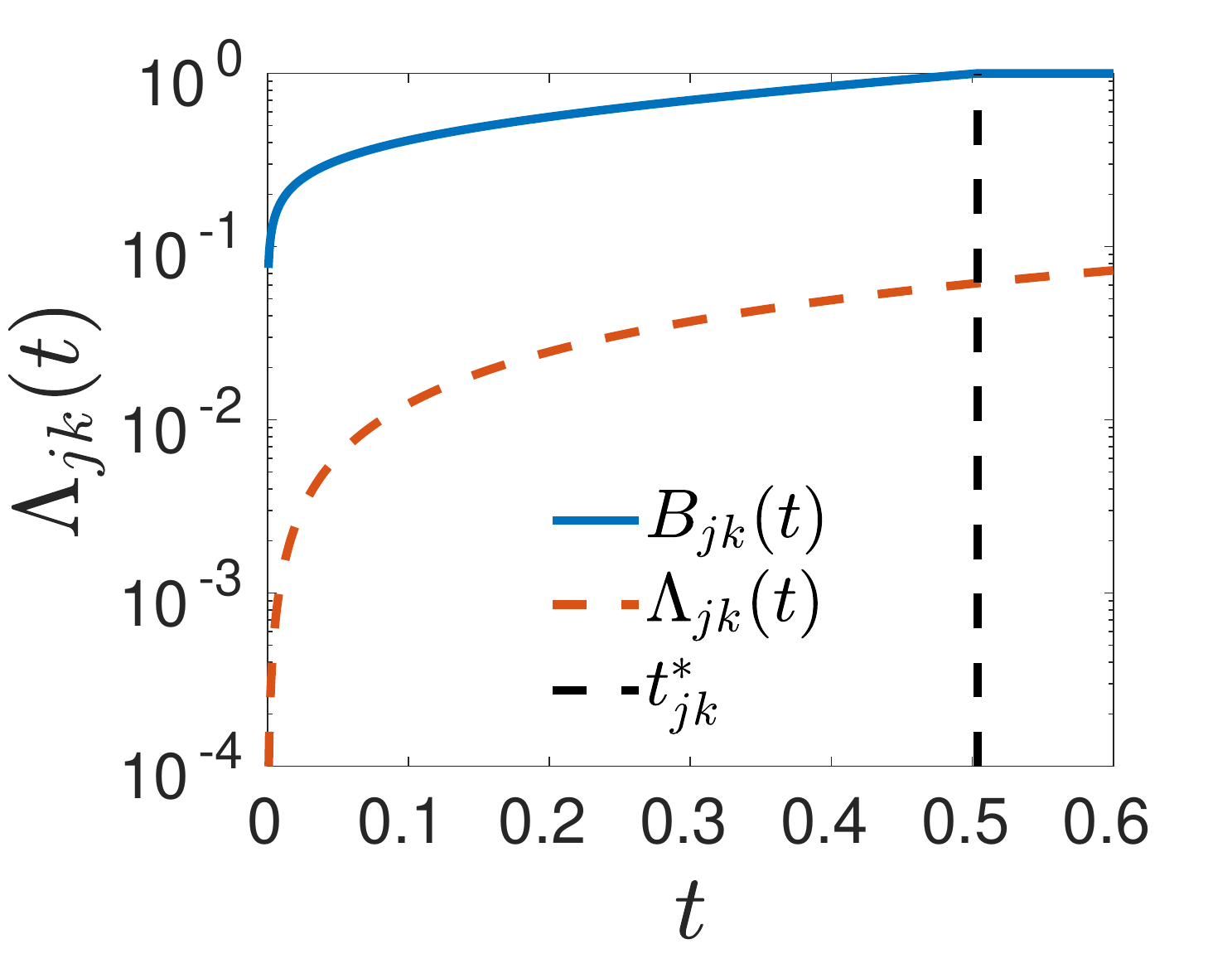}%
	\hspace{-8pt}%
	\includegraphics[width=0.53\linewidth]{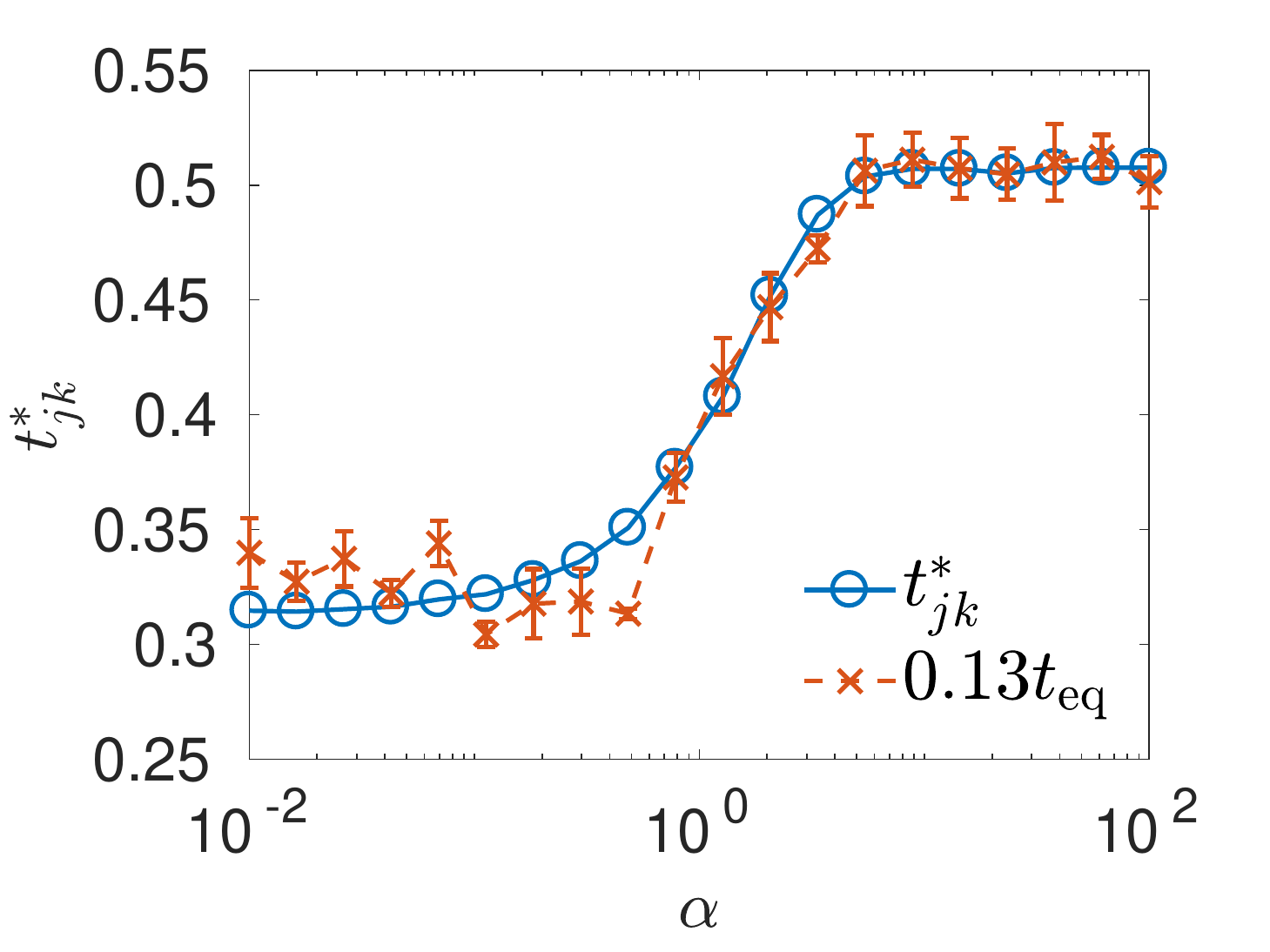}
	\caption{Comparison of our bounds with exact results for the transverse-field Ising model \eqref{eq:ex_H} on $L=8$ lattice sites. Left: The bound $B_{jk}$ (solid line) compared to $\varLambda_{jk}$  (dashed line) as a function of time for $\alpha=10$. Right: Comparison of the timescale $t^\ast_{jk}$ and the estimated equilibration time $t_\text{eq}$, determined according to the protocol described in the main text, as a function of $\alpha$. Error bars indicate standard errors resulting from averaging over 10 randomly chosen nonequilibrium initial states.} \label{fig:LR_evol}
\end{figure}

Equation \eqref{eq:LR_bd3} gives an upper bound on how strongly the population of the observable eigenstate $\kt{O_k}$, encoded in the variable $x_k$, can influence the population of $\kt{O_j}$ at a later time $t$; see Fig.~\ref{fig:LR} for an illustration. The influence of $x_k$ on $x_j$ is small initially, and it grows with increasing $t$. The larger the difference $|O_k-O_j|$, the longer it takes for $\varLambda_{jk}$ to become non-negligible. Thinking of $O_k$ as belonging to the nonequilibrium subspace $\H_\text{neq}$ and of $O_j$ as belonging to $\H_\text{eq}$, the time $t^\ast_{jk}$ in Eq.~\eqref{eq:tstar} gives a lower bound on the equilibration timescale because it separates the light-cone-shaped causal region where $B_{jk}=0$ from the region where $B_{jk}<0$. In the latter region, the influence of $\H_\text{neq}$ on $\H_\text{eq}$ is negligible and the system cannot have equilibrated yet. We therefore take $t^\ast_{jk}$ as a lower estimate of the equilibration timescale, $t_\text{eq}\geq t^\ast_{jk}$. The slope of the light cone can be read off from Eq.~\eqref{eq:tstar} and estimates the speed of oscillations propagating through the network. 

The bound $B_{jk}(t)$ and the timescale $t^\ast_{jk}$ depend on the product of parameters $\kappa gr$, where $\kappa$ and $r$ are affected by the locality of the observable $O$, and $g$ is affected by the locality of $H$. To illustrate this dependence, we fix $\|H\|=k=r=1$ and consider the maximum distance $|O_k-O_j|=2$ away from equilibrium. For a strongly local Hamiltonian with only nearest-neighbor pair interactions we have $g\sim\|H\|/L$, implying
\begin{equation}
t^\ast_{jk} = \frac{L}{L-1} W(\frac{L-1}{e})\sim \ln(L)
\end{equation}
in the large-system limit. In the absence of any locality, we can assume $g\sim\|H\|(L-1)/L$, which results in the scaling
\begin{equation}
t^\ast_{jk} = W\left(\frac{1}{e(L-1)}\right)L\sim \frac{1}{e};
\end{equation}
see Fig.~\ref{fig:graph_tstL} for an illustration. This dependence of equilibration times on locality is in qualitative agreement with findings for specific models \cite{vdWorm_etal13,KastnerVdWorm15,EisertvdWormManmanaKastner13,Mori19}. Unlike other Lieb-Robinson bounds, the right-hand side of Eq.~\eqref{eq:LR_bd3} is not uniform in the system size $L$ but grows with $L$ through $\|H\|$, and it possibly also grows through system-size dependencies of $\kappa$, $g$, and $r$. 

{\em Transverse-field Ising model.---}When drawing conclusions based on a bound, it is instructive to investigate the tightness of the bound by comparing to exact results. We consider the Hamiltonian
\begin{align}
 H 
 = \sum_{i=1}^{L-1} \sum_{j=i+1}^L h_{ij} + \frac{\Gamma}{Z}\sum_{i=1}^{L}\s^z_i \label{eq:ex_H}
\end{align}
of a spin chain with open boundary conditions and pair interactions
\begin{align}\label{eq:pair}
  h_{ij}=\frac{1}{Z}\frac{J}{|i-j|^\a} \s^z_i \s^z_j,
\end{align}
where $\sigma^z_i$ and $\sigma^x_i$ denote the $z$ and $x$ components of a Pauli spin operator acting on lattice site $i$. We consider a coupling coefficient of $J=1$ and set the external field to $\Gamma=5$. The coupling strength decays with the distance $|i-j|$ between lattice sites like a power law with exponent $\a$. The larger $\alpha$, the more local the interactions and the smaller the locality parameter $g$. The normalization constant $Z$ is chosen such that $\|H\|=1$. This guarantees that, upon variation of $\alpha$, the speed of equilibration is affected only by a change of locality but not trivially by a change of the norm of $H$.

We study equilibration of the magnetization 
\begin{align}\label{eq:ex_O}
   M = \frac{1}{L} \sum_{i=1}^L \s^z_i,
\end{align}
for which the locality parameters take on the values $\kappa=1$ and $r=1$. The eigenstates $\kt{M_j}$ of $M$ are products of eigenstates of $\s^z$, with eigenvalues $M_j\in\{-L,-L+2,\dotsc,L-2,L\}$. For almost all initial states, the equilibrium eigenstates of $M$ correspond to eigenvalues $M_j\approx0$.
Figure~\ref{fig:LR_evol} (left) compares numerical results for the time evolution of $\varLambda_{jk}$, which is obtained by exact diagonalization for a chain of $L=8$ spins, to the bound $B_{jk}$ \eqref{eq:LR_bd3}. As is common for Lieb-Robinson-type bounds, $B_{jk}$ overestimates $\varLambda_{jk}$ substantially. The functional form of the initial increase of the exact $\varLambda_{jk}$, however, is well captured by $B_{jk}$. The timescale $t^\ast_{jk}$ marks the end of the rapid increase of $\varLambda_{jk}(t)$, confirming the use of $t^\ast_{jk}$ as a lower bound on $t_\text{eq}$.

To further compare $t^\ast_{jk}$ and $t_\text{eq}$, we estimate $t_\text{eq}$ for 10 random initial states with fixed amplitude $|x_k(0)|^2=0.8$, where $\kt{M_k}$ is the nonequilibrium eigenstate with eigenvalue $M_k=1$ maximizing the distance to the equilibrium value $M_j\approx0$. The estimation of $t_\text{eq}$ is done by finding the earliest time where the variance $\bropkt{\psi(t)}{(M-\langle M\rangle_\text{eq})^2}{\psi(t)}$ drops below 10\% of its longtime average, which is a procedure that captures the essence of our definition of equilibrium further above. In Fig.~\ref{fig:LR_evol} (right), we compare $t^\ast_{jk}$ and the numerically estimated $t_\text{eq}$ for various $\alpha$. For a better comparison of the functional dependencies, $t_\text{eq}$ has been rescaled by a factor of $0.13$. The results confirm $t_\text{eq}\geqslant t^\ast_{jk}$ for all $\alpha$, as well as the expected increase of the equilibration timescale with increasing $\alpha$. Moreover, the estimate $t^\ast_{jk}$ captures the functional form of the $\alpha$ dependence of $t_\text{eq}$ remarkably well.

We also compared our bounds to the exact dynamics of disordered systems and systems with nonalgebraic decay of coupling strength (not shown). In all examples, the validity of the bounds $B_{jk}$ and $t^\ast_{jk}$ is confirmed, the initial increase of $B_{jk}(t)$ captures the functional form of $\varLambda_{jk}(t)$ to the same extent as in Fig.~\ref{fig:LR_evol}, and (except for specific choices of the parameters) the $\alpha$ dependence of $t^\ast_{jk}$ agrees qualitatively with that of the measured $t_\text{eq}$.

{\em Sketch of the proof of Eq.~\eqref{eq:LR_bd3}.---}Starting from the bound on the right-hand side of Eq.~\eqref{eq:LR_obj2}, we  adapt a strategy 
used by Arad {\em et al.}\ \cite{AradKuwaharaLandau16} and de Oliveira {\em et al.}\ \cite{deOliveira_etal18} to derive a bound on the matrix elements of a local observable in the eigenbasis of a local Hamiltonian. Introducing the auxiliary variable $s\geqslant0$, we use $e^{-sO}e^{sO}=1$ and write the Hamiltonian as 
$(H^n)_{jk} = \bropkt{O_j}{e^{sO}\,H^n\,e^{-sO}}{O_k} \, e^{-s|o_k-o_j|}$.
Using Hadamard's formula
\begin{equation}
e^{sO}\,H^n\,e^{-sO} = \sum_{l=0}^\infty \frac{s^l}{l!} K_l^{(n)}
\end{equation}
with the $l$-nested commutator $K_l^{(n)} = [O,\dots,[O,H^n]\cdots]$, we obtain
\begin{equation}\label{eq:LR_bd2}
	\varLambda_{jk}(t) \leqslant e^{-s|o_k-o_j|} \sum_{n=0}^\infty  \frac{t^n}{n!} \sum_{l=0}^\infty \frac{s^l}{l!} \|K_l^{(n)}\|.
\end{equation}
Writing the Hamiltonian and the observable as sums over local terms, most of the local commutators in $K_l^{(n)}$ vanish and, by making use of combinatorial techniques detailed in the Supplemental Material, we obtain
\begin{equation}\label{eq:l_comm_bd}
	\|K_l^{(n)}\| \leqslant \sum_{j=1}^{\min(l,n)} \sn{l}{j}\frac{n!}{(n-j)!}\, (2\|O\|)^l\,(kgr)^j\,\|H\|^{n-j},
\end{equation}
where $\{\substack{l\\j}\}$ denotes Stirling numbers of the second kind. After some algebra detailed in the Supplemental Material, we obtain
\begin{align}\label{eq:s_bound}
	\ln\varLambda_{jk}(t) \leqslant \left(\|H\| - kgr\right)t + kgrt \, e^{2s\|O\|} - s|o_k-o_j|.
\end{align}
Minimizing the right-hand side of Eq.~\eqref{eq:s_bound} over $s\geqslant0$ we arrive at our main result \eqref{eq:LR_bd3}.

{\em Conclusions.---}By rewriting a quantum system as a classical network on Hilbert space, we derived a Lieb-Robinson-type upper bound on the spreading of a perturbation across Hilbert space. Based on this rigorous result \eqref{eq:LR_bd3}, we provided a lower estimate of the equilibration time of the corresponding quantum system. On the technical side, the progress reported in our work is the result of a twofold change of viewpoint: firstly the mentioned interpretation of a quantum system as a classical network in Hilbert space, to which classical Lieb-Robinson techniques may be applied; and secondly, different from existing results in the literature, the focus on a {\em lower}\/ bound on the equilibration time. 

On the conceptual side, the main novelty of our work is that the degree of locality of the Hamiltonian and observable enters the bound \eqref{eq:LR_bd3}. Quantified through the parameters $\kappa$, $g$, and $r$, locality is believed to be a key player, which is crucial in determining the equilibration time of a quantum system. In the language of a classical network in Hilbert space, locality implies sparseness of the network, which in turn reduces the speed at which a perturbation can travel across the network. Indeed, in the case of pronounced locality, our bound predicts that the timescale $t^\ast_{jk}$ in Eq.~ \eqref{eq:tstar} scales doubly logarithmic with the dimension of the Hilbert space. Our results are confirmed by exact diagonalization for small system sizes, where $t^\ast_{jk}$ is not only found to lower bound the observed equilibration times but also qualitatively captures some of their functional dependencies. The rather weak notion of locality \eqref{eq:g} that we made use of must be considered as a first step towards physically realistic estimates. Refinements that employ locality in a stronger sense are a promising direction for future research.

Despite these evident successes, it is worth emphasizing that the generality of our results necessarily implies that the bounds, albeit valid, cannot be tight in all cases. Although the actual equilibration time of the quantum system is expected to depend on the specific observable and initial state considered, the choice of the observable enters in our bound only through the locality parameters $\kappa$ and $r$, and the choice of the initial state enters only through $|O_k-O_j|$. Considering two specific nodes $x_i$ and $x_j$ of the classical network such that $O_i=O_j$, it may be the case that one of the two is less strongly connected to the rest of the network, and accordingly equilibrates more slowly, whereas our bound estimates the corresponding equilibration timescales of $x_i$ and $x_j$ to be identical.

\acknowledgments
M.\,K.\ acknowledges financial support by the South African National Research Foundation through the Incentive Funding Programme and the Competitive Funding for Rated Researchers.

\bibliography{MK}

\newpage

\includepdf[pages={1},scale=1,pagecommand={\thispagestyle{empty}}]{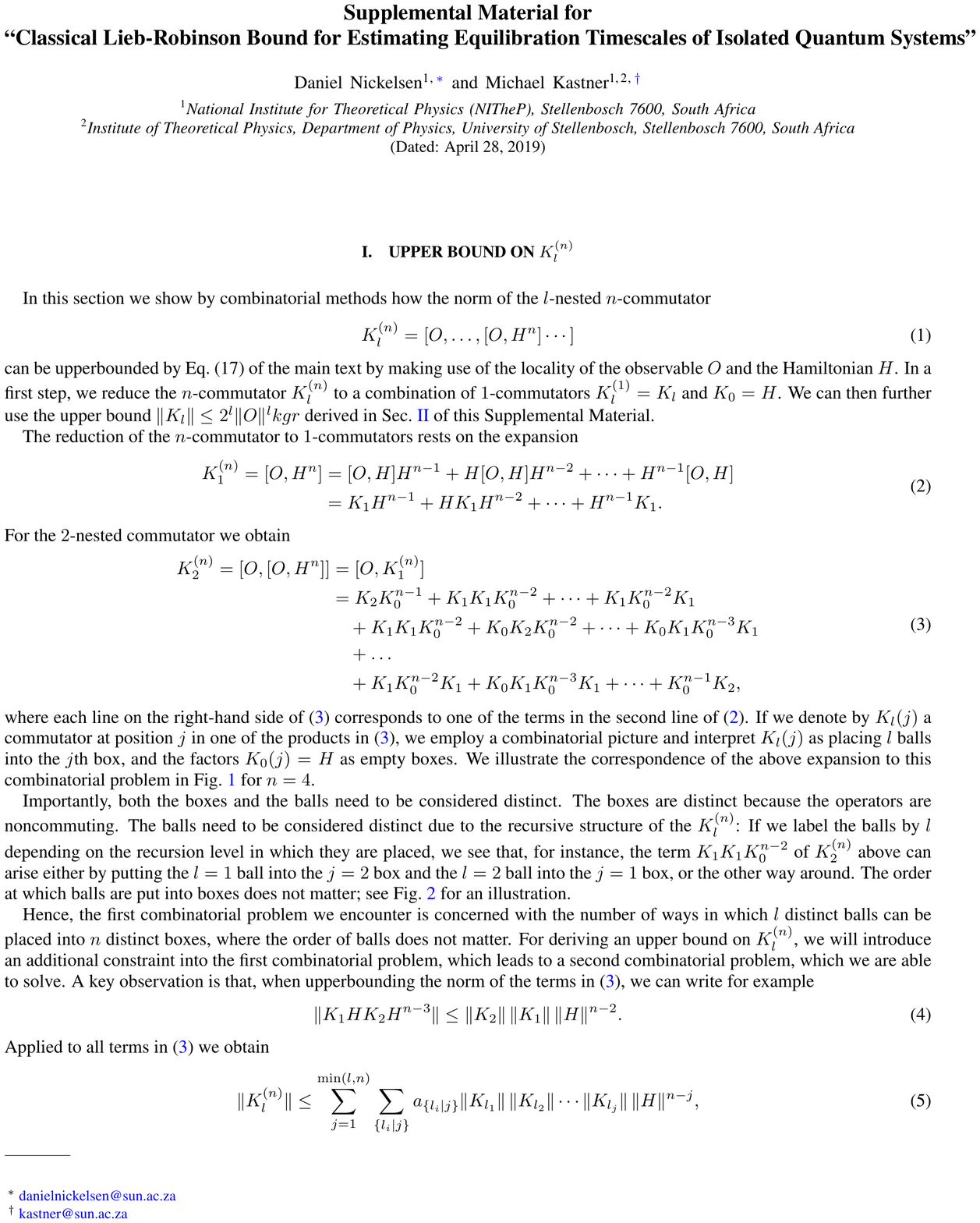}
\hbox{}\newpage
\includepdf[pages=2]{./EquiLocalSM.pdf}
\hbox{}\newpage
\includepdf[pages=3]{./EquiLocalSM.pdf}
\hbox{}\newpage


\section*{Supplementary material}

Daniel Nickelsen[1, *] and Michael Kastner[1, 2, †]

[1]*National Institute for Theoretical Physics (NITheP), Stellenbosch 7600, South Africa*
[2]*Institute of Theoretical Physics, Department of Physics, University of Stellenbosch, Stellenbosch 7600, South Africa*
(Dated: April 28, 2019)

## I. UPPER BOUND ON $K_l^{(n)}$

In this section we show by combinatorial methods how the norm of the $l$-nested $n$-commutator

$$K_l^{(n)} = [O, \ldots, [O, H^n] \cdots] \tag{1}$$

can be upperbounded by Eq. (17) of the main text by making use of the locality of the observable $O$ and the Hamiltonian $H$. In a first step, we reduce the $n$-commutator $K_l^{(n)}$ to a combination of 1-commutators $K_l^{(1)} = K_l$ and $K_0 = H$. We can then further use the upper bound $\|K_l\| \leq 2^l \|O\|^l kgr$ derived in Sec. II of this Supplemental Material.

The reduction of the $n$-commutator to 1-commutators rests on the expansion

$$\begin{aligned}
K_1^{(n)} = [O, H^n] &= [O, H]H^{n-1} + H[O, H]H^{n-2} + \cdots + H^{n-1}[O, H] \\
&= K_1 H^{n-1} + H K_1 H^{n-2} + \cdots + H^{n-1} K_1.
\end{aligned} \tag{2}$$

For the 2-nested commutator we obtain

$$\begin{aligned}
K_2^{(n)} = [O, [O, H^n]] &= [O, K_1^{(n)}] \\
&= K_2 K_0^{n-1} + K_1 K_1 K_0^{n-2} + \cdots + K_1 K_0^{n-2} K_1 \\
&\quad + K_1 K_1 K_0^{n-2} + K_0 K_2 K_0^{n-2} + \cdots + K_0 K_1 K_0^{n-3} K_1 \\
&\quad + \ldots \\
&\quad + K_1 K_0^{n-2} K_1 + K_0 K_1 K_0^{n-3} K_1 + \cdots + K_0^{n-1} K_2,
\end{aligned} \tag{3}$$

where each line on the right-hand side of (3) corresponds to one of the terms in the second line of (2). If we denote by $K_l(j)$ a commutator at position $j$ in one of the products in (3), we employ a combinatorial picture and interpret $K_l(j)$ as placing $l$ balls into the $j$th box, and the factors $K_0(j) = H$ as empty boxes. We illustrate the correspondence of the above expansion to this combinatorial problem in Fig. 1 for $n = 4$.

Importantly, both the boxes and the balls need to be considered distinct. The boxes are distinct because the operators are noncommuting. The balls need to be considered distinct due to the recursive structure of the $K_l^{(n)}$: If we label the balls by $l$ depending on the recursion level in which they are placed, we see that, for instance, the term $K_1 K_1 K_0^{n-2}$ of $K_2^{(n)}$ above can arise either by putting the $l = 1$ ball into the $j = 2$ box and the $l = 2$ ball into the $j = 1$ box, or the other way around. The order at which balls are put into boxes does not matter; see Fig. 2 for an illustration.

Hence, the first combinatorial problem we encounter is concerned with the number of ways in which $l$ distinct balls can be placed into $n$ distinct boxes, where the order of balls does not matter. For deriving an upper bound on $K_l^{(n)}$, we will introduce an additional constraint into the first combinatorial problem, which leads to a second combinatorial problem, which we are able to solve. A key observation is that, when upperbounding the norm of the terms in (3), we can write for example

$$\|K_1 H K_2 H^{n-3}\| \leq \|K_2\| \|K_1\| \|H\|^{n-2}. \tag{4}$$

Applied to all terms in (3) we obtain

$$\|K_l^{(n)}\| \leq \sum_{j=1}^{\min(l,n)} \sum_{\{l_i|j\}} a_{\{l_i|j\}} \|K_{l_1}\| \|K_{l_2}\| \cdots \|K_{l_j}\| \|H\|^{n-j}, \tag{5}$$

---
* danielnickelsen@sun.ac.za
† kastner@sun.ac.za

$$K_2^{(4)} =$$

FIG. 1. The first combinatorial problem. Shown is a graphical representation of the expansion of $K_2^{(4)}$ as an example for (3). The distinct balls are enumerated according to $l = 1$ and $l = 2$.

where the second sum is over all integer partitions of $l$ into $j$ parts (i.e., all sets of $j$ nonnegative integers $\{l_i\}$ satisfying $l_1 + l_2 + \cdots + l_j = l$) and the first sum is over all possible numbers of non-empty boxes, which is limited by $\min(l, n)$ from above. Since the factors in each term now commute, terms that were distinct before are now identical, and we need to account for the multiplicities of those terms. In (5), we denoted these multiplicities by $a_{\{l_i|j\}}$, to be determined later.

At this point we substitute the upper bound (21), derived in Sec. II of this Supplemental Material, into (5) to obtain

$$\sum_{\{l_i|j\}} a_{\{l_i|j\}} \|K_{l_1}\| \|K_{l_2}\| \cdots \|K_{l_j}\| \|H\|^{n-j} \le p_j(n,l)(2\|O\|)^l (kgr)^j \|H\|^{n-j}. \tag{6}$$

The terms on the right-hand side of (6) do not anymore depend on the individual partitions $\{l_i|j\}$, but only on the number of terms that correspond to $j$ non-empty boxes. The second combinatorial problem hence asks for the number of ways in which $l$ distinct balls can be placed into $n$ distinct boxes under the constraint that only $j$ of the $n$ boxes are used. The answer to this problem determines the factor $p_j(n,l) \equiv \sum_{\{l_i|j\}} a_{\{l_i|j\}}$ in (6), which can be derived in two steps.

First we use that the Stirling number of second kind determines the number of ways in which $l$ distinct balls can be placed into $j \le n$ *identical* boxes, under the additional requirement that no box remains empty (and the order of balls do not matter),

$$S(l,j) \equiv \left\{ \begin{matrix} l \\ j \end{matrix} \right\} = \frac{1}{j!} \sum_{k=0}^{j} (-1)^{j-k} \binom{j}{k} k^l. \tag{7}$$

In the second step, we need to factor in the extra configurations due to $n - j$ empty boxes and having distinct boxes instead of identical boxes. In Fig. 3 we illustrate the situation for $l = 6$ balls and $n = 4$ boxes for different numbers $j$ of non-empty boxes. To account for the extra configurations, we need to swap the boxes for each $j$ in all possible ways. The number of permutations over all $n$ boxes, including the empty ones, is $n!/(n-j)!$, which solves our second combinatorial problem

$$p_j(n,l) = \left\{ \begin{matrix} l \\ j \end{matrix} \right\} \frac{n!}{(n-j)!}. \tag{8}$$

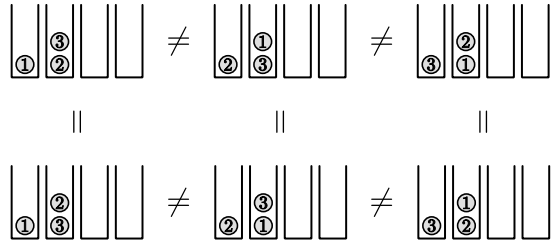

FIG. 2. Illustration of the configurations that need to be considered distinct in order to correctly count the terms in $K_l^{(n)}$, here for $K_3^{(4)}$. The configurations in the top and bottom rows are distinct due to the balls being distinct, whereas the configurations in the columns are identical since the order of balls does not matter. The balls are labelled according to $l = 1$, $l = 2$ and $l = 3$.

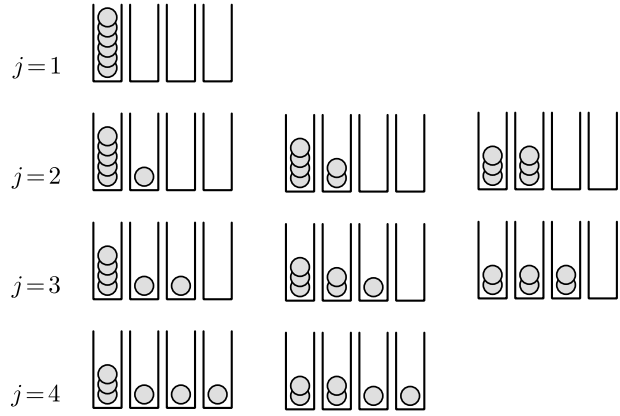

FIG. 3. The second combinatorial problem. Each row considers the number of ways to place $l = 6$ distinct balls into $j \leq n = 4$ *identical* boxes, under the additional constraint that no box remains empty (permutations due to distinct balls are not shown; order of balls do not matter, c.f. Fig. 2). The number of configurations of each $j$ is given by the Stirling number of the second kind, $S(l, j)$ in (7), the number of permutations due to all $n$ *distinct* boxes, including the $n - j$ empty boxes, is given by the factor $n!/(n-j)!$.

Our reasoning is confirmed by the sum rule for $l \leq n$,

$$\sum_{j=1}^{l} \left\{ \begin{matrix} l \\ j \end{matrix} \right\} \frac{n!}{(n-j)!} = n^l, \tag{9}$$

which agrees with the total number of terms in the expansion of $K_l^{(n)}$ in the first combinatorial problem.

Plugging (6) and (8) into (5) we arrive at the bound

$$\|K_l^{(n)}\| \leq (2\|O\|)^l \sum_{j=1}^{\min(l,n)} \left\{ \begin{matrix} l \\ j \end{matrix} \right\} \frac{n!}{(n-j)!}(kgr)^j \|H\|^{n-j}. \tag{10}$$

Inserting (10) into Eq. (16) of the main text, we can write

$$\Lambda_j k(t) \leq \sum_{n=0}^{\infty} \frac{t^n}{n!}|(H^n)_{jk}| \leq \sum_{n=0}^{\infty} \frac{t^n}{n!}e^{-s|o_j - o_k|}\sum_{l=0}^{\infty} \frac{s^l}{l!}\|K_l^{(n)}\| \tag{11a}$$

$$\leq e^{-s|o_j - o_k|}\sum_{n=0}^{\infty} \frac{t^n}{n!}\sum_{l=0}^{\infty} \frac{s^l}{l!}(2\|O\|)^l \sum_{j=1}^{\min(l,n)} \left\{ \begin{matrix} l \\ j \end{matrix} \right\} \frac{n!}{(n-j)!}(kgr)^j \|H\|^{n-j} \tag{11b}$$

$$= e^{-s|o_j - o_k|}\sum_{n=0}^{\infty} t^n \sum_{l=0}^{\infty} \frac{(2s\|O\|)^l}{l!} \sum_{j=1}^{\min(l,n)} \left\{ \begin{matrix} l \\ j \end{matrix} \right\} \frac{\|H\|^{n-j}}{(n-j)!}(kgr)^j. \tag{11c}$$

Next we exchange the order of the sums and split the $n$-sum into contributions for $n \leq l$ and $n > l$, respectively, leading to

$$\sum_{n=0}^{\infty} t^n \sum_{l=0}^{\infty} \frac{(2s\|O\|)^l}{l!} \sum_{j=1}^{\min(l,n)} \left\{ \begin{matrix} l \\ j \end{matrix} \right\} \frac{\|H\|^{n-j}}{(n-j)!}(kgr)^j = \sum_{l=0}^{\infty} \frac{(2s\|O\|)^l}{l!}\Big[\Sigma_1(l) + \Sigma_2(l)\Big] \tag{12}$$

with

$$\Sigma_1 = \sum_{n=0}^{l} t^n \sum_{j=1}^{n} \left\{ \begin{matrix} l \\ j \end{matrix} \right\} \frac{\|H\|^{n-j}}{(n-j)!}(kgr)^j = \sum_{j=1}^{l} \left\{ \begin{matrix} l \\ j \end{matrix} \right\} (kgr)^j \sum_{n=j}^{l} t^n \frac{\|H\|^{n-j}}{(n-j)!} \tag{13}$$

and

$$\Sigma_2 = \sum_{n=l+1}^{\infty} t^n \sum_{j=1}^{l} \left\{ \begin{matrix} l \\ j \end{matrix} \right\} \frac{\|H\|^{n-j}}{(n-j)!}(kgr)^j = \sum_{j=1}^{l} \left\{ \begin{matrix} l \\ j \end{matrix} \right\} (kgr)^j \sum_{n=l+1}^{\infty} t^n \frac{\|H\|^{n-j}}{(n-j)!}. \tag{14}$$

We can then recombine $\Sigma_1$ and $\Sigma_2$ and simplify to

$$\sum_{l=0}^{\infty} \frac{(2s\|O\|)^l}{l!}\Big[\Sigma_1(l) + \Sigma_2(l)\Big] = \sum_{l=0}^{\infty} \frac{(2s\|O\|)^l}{l!} \sum_{j=1}^{l} \left\{ \begin{matrix} l \\ j \end{matrix} \right\} (kgr)^j \sum_{n=j}^{\infty} t^n \frac{\|H\|^{n-j}}{(n-j)!} \tag{15a}$$

$$= \sum_{l=0}^{\infty} \frac{(2s\|O\|)^l}{l!} \sum_{j=1}^{l} \left\{ \begin{matrix} l \\ j \end{matrix} \right\} (kgr)^j e^{\|H\|t} t^j \tag{15b}$$

$$= e^{\|H\|t} \sum_{l=0}^{\infty} \frac{(2s\|O\|)^l}{l!} \sum_{j=1}^{l} \left\{ \begin{matrix} l \\ j \end{matrix} \right\} z^j = e^{\|H\|t} \sum_{l=0}^{\infty} \frac{(2s\|O\|)^l}{l!} \mathrm{B}_l(z) \tag{15c}$$

$$= e^{\|H\|t} \exp\Big[\Big(e^{2s\|O\|} - 1\Big)z\Big] \tag{15d}$$

with $z = kgrt$. $\mathrm{B}_l(z)$ in (15c) denotes the Bell polynomials, whose exponential generating function is known and leads to (15d). Plugging (12) and (15d) into (11c) we obtain Eq. (18) of the main text.

## II. UPPER BOUND ON $K_l$

In this section we adapt a strategy from Refs. [1, 2] to obtain, for local observables and Hamiltonians, an upper bound on the $l$-nested commutator

$$K_l = [O, \ldots, [O, H] \cdots]. \tag{16}$$

We estimate $\|K_l\|$ in terms of $\|K_1\|$ by writing

$$\|K_l\| \leq \|[O, \ldots, [O, H] \cdots]\| \leq 2^{l-1}\|O\|^{l-1}\|K_1\|. \tag{17}$$

Making use of the decompositions of Hamiltonian and observable as sums of local terms as introduced in the main text, the 1-nested commutator becomes

$$K_1 = \sum_{X,Y} [\omega_Y, h_X]. \tag{18}$$

At the expense of doublecounting of some terms, we can rewrite the summations in (18) as

$$\|K_1\| \leq \Big\| \sum_{Y} \sum_{i \in Y} \sum_{X \ni i} [\omega_Y, h_X] \Big\|, \tag{19}$$

where $X \ni i$ denotes all subsets $X \in \mathscr{N}$ that contain $i$. For a $\kappa$-local observable as introduced in Eq. (2) of the main text, we can upperbound

$$\|K_1\| \leq \sum_{y} k \Big\| \sum_{x \ni i} [\omega_y, h_x] \Big\| \leq \sum_{y} 2k\|\omega_y\| \Big\| \sum_{x \ni i} h_x \Big\| \leq 2\|O\|kgr \tag{20}$$

with $g$ and $r$ as defined in the main text. Inserting (20) into (17) we obtain the bound

$$\|K_l\| \leq 2^l \|O\|^l kgr \tag{21}$$

for the $l$-nested commutator.

The above strategy of proof could also be applied directly to the $l$-nested commutator (16), without resorting to (17), by iteratively counting the number of nonvanishing terms in $[O, K_1], [O, K_2], \ldots, [O, K_{l-1}]$. Surprisingly, because of excessive multiple-counting of terms, the resulting bound is less tight than (21) and entails divergences in the subsequent calculations.

---



\end{document}